\documentclass[aip,jcp,reprint]{revtex4-1}

\usepackage{amsmath}
\usepackage{amssymb}
\usepackage{mathtools}
\usepackage{graphicx}
\usepackage{hhline}

\usepackage{color}
\DeclareGraphicsExtensions{.eps}

\makeatletter
\def\fixedlabel#1#2{%
  \@bsphack%
  \protected@write\@auxout{}%
         {\string\newlabel{#1}{{#2}{\thepage}}}%
  \@esphack}
\makeatother

\begin{document}


\title{Vapour-Liquid Coexistence of an Active Lennard-Jones fluid} 

\author{Vasileios Prymidis }
\email[\noindent Electronic mail: ]{v.prymidis@uu.nl}
\noaffiliation

\author{\normalfont\textsuperscript{b)}}
\noaffiliation

\author{Siddharth Paliwal}
\thanks{\noindent These authors have contributed equally to this work.}
\noaffiliation

\author{Marjolein Dijkstra}
\noaffiliation

\author{Laura Filion}
\noaffiliation

\affiliation{Soft Condensed Matter, Debye Institute for Nanomaterials Science, Utrecht University, Princetonplein 5, 3584 CC Utrecht, The Netherlands}

\date{\today}
%
%
%
%
\begin{abstract}

We study a three-dimensional system of self-propelled Lennard-Jones particles using Brownian Dynamics simulations. Using recent theoretical results for active matter, we calculate the pressure and report equations of state for the system. Additionally, we chart the vapour-liquid coexistence and show that the coexistence densities can be well described using simple power laws. Lastly, we demonstrate that our out-of-equilibrium system shows deviations from both the law of rectilinear diameters and the law of corresponding states.
\end{abstract}


\maketitle 

\section{Introduction}

%
%
%
%
\label{Introduction}

Active matter has recently emerged as an important paradigm for out-of-equilibrium systems.\cite{ramaswamy2010mechanics, romanczuk2012active, marchetti2013hydrodynamics} Experimental breakthroughs in the fabrication and observation of colloidal swimmers,\cite{paxton2004catalytic,dreyfus2005microscopic,howse2007self,hong2007chemotaxis,deseigne2010collective,kudrolli2008swarming,schaller2010polar,volpe2011microswimmers,palacci2013living,bricard2013emergence}  have inspired a boom  of theoretical studies of self-propelled particles in soft matter physics. In particular, there has been considerable interest in exploring the applicability of equilibrium statistical physics concepts, such as pressure and surface tension, to describe active matter and associated phase transitions.\cite{wittkowski2014scalar, takatori2015towards, maggi2015multidimensional, solon2015pressure, bialke2015negative, falasco2015mesoscopic, speck2015ideal, speck2016stochastic} 
One of the few systems where a phase transition has been thoroughly explored, even in the context of critical phenomena, is the Vicsek model together with its modifications.\cite{vicsek1995novel, chate2006simple, chate2008collective, ginelli2010relevance, dey2012spatial} In this work, we study a different, yet also highly important model system for active matter, namely a system of attractive isotropic self-propelled, Brownian particles.\cite{romanczuk2012active, bechinger2016active} We investigate theoretically  a vapour-liquid phase transition and present an extensive study of the out-of-equilibrium phase transition.  

One of the most well-studied equilibrium model systems which undergo such a vapour-liquid phase transition is the Lennard-Jones (LJ) fluid.\cite{smit1992phase, vrabec2006comprehensive}  In this paper, we modify this model by introducing a self-propulsion force to each particle, and treat the motion using Brownian dynamics.  This model was chosen since the phase behaviour of the equilibrium system is well characterized and can be readily verified by computer simulations. In particular, since the LJ particles interact via a short-range attractive potential, the second-order vapour-liquid phase transition of the system belongs to the Ising universality class.\cite{yang1952statistical, lee1952statistical2, watanabe2012phase} Moreover, the LJ fluid obeys both the law of rectilinear diameters when in phase coexistence, which is obeyed by a myriad of real substances as well as active matter systems,\cite{watanabe2012phase, reif2010history, prymidis2016state} and the Noro-Frenkel law of corresponding states, which maps the thermodynamic properties of different spherically symmetric attractive potentials onto each other.\cite{noro2000extended, dunikov2001corresponding} 

In the case of the active LJ fluid preliminary studies of the vapour-liquid phase transition have hinted on the deviation of the transition properties from equilibrium.\cite{redner2013reentrant, Mognetti2013, prymidis2015self} Most interestingly, when the  direction of the self-propulsion of the particles diffuses in  a sufficiently slow rate, a percolating state was found between the fluid and the vapour-liquid coexistence. \cite{prymidis2015self}  Herein we expand on these results by measuring the equation of state for the system and studying the behaviour of the pressure in the vapour-liquid coexistence regime. Additionally, we map out the phase diagram for different combinations of the propulsion speed and rotational diffusion rate. We compare the behaviour of the binodals of the active system with that of the equilibrium system by exploring whether the laws of rectilinear diameters and corresponding states hold. Moreover, we examine whether the binodals can be fitted via simple power laws. 
 
In section \ref{Model} we introduce the model and the dynamics and also present the method that we used to calculate the equation of state. Equations of state  are presented in section \ref{equationsofstate}, followed by a close study of the phase coexistence in section \ref{gasliquidcoexistence}. This study includes the power law and exponential scaling of the order parameter and the critical temperature respectively in section \ref{orderparameter}, a test of the law of rectilinear diameters and the scaling of the critical density in section \ref{lawofrectilinear} and ultimately a test of the law of corresponding states in  section \ref{binodals}. Our conclusions are summarized in section \ref{Discussion}.

%
%
%
%
\section{Methods}
\label{Model}

\subsection{Model and Dynamics}

 We consider a three-dimensional system consisting of self-propelled spherical particles (colloids) immersed in a molecular solvent, in a periodic  box with dimensions $L_x, L_y$ and $L_z$. The position of the center of mass of the $i$th particle  at time $t$ is given by the vector $\boldsymbol{r}_i(t)$. With particle $i$, we associate a three-dimensional unit vector $\boldsymbol{u}_i(t)$  that indicates the direction of the self-propelling force. The particles interact with each other  via a Lennard-Jones potential 
\begin{equation}
   U({r}_{ij})=  
  4\epsilon \left[ \left( \frac{\sigma}{{r}_{ij}}\right)^{12}-\left( \frac{\sigma}{{r}_{ij}}\right)^6
  \right] ,
 \end{equation}
truncated and shifted at $2.5 \sigma$, where $\sigma$ is  the particle length scale, ${r}_{ij}=|\boldsymbol{r}_{j}-\boldsymbol{r}_{i}|$ and the parameter $\epsilon$ controls the strength of the interaction. 

To describe the translational and rotational motion of the individual colloidal particles inside the solvent we use the overdamped Langevin equations
\begin{align}\label{langevin}
 \frac{d\boldsymbol{r}_i}{dt}& =
 -\frac{1}{\eta}\sum_{j\neq i}\frac{\partial U(r_{ij})}{\partial \boldsymbol{r}_i}   + \upsilon_0 \boldsymbol{u}_i + \sqrt{2D_{tr}}\boldsymbol{\xi}^{tr}_{i},
 \\
 \label{roteq}
  \frac{d\boldsymbol{u}_i}{dt} &=  \sqrt{2D_r}\left(\boldsymbol{u}_i \times \boldsymbol{\xi}^{r}_i\right).
 \end{align}
The translational diffusion coefficient is  given by the Einstein-Smoluchowski relation $ D_{tr}=1/(\beta_s \eta) $, with $\eta$ the damping coefficient and $\beta_s$ the inverse temperature of the surrounding solvent. $D_r$ denotes the rotational diffusion coefficient and $\upsilon_0$  the propulsion speed. The vectors  $\boldsymbol{\xi}^{tr}_i $ and $\boldsymbol{\xi}^{r}_i $ are  unit-variance random vectors, with mean value and variation 
\begin{align}\label{variance1}
  \langle \boldsymbol{\xi}^{tr,r}_i(t) \rangle &=0,
\\
  \langle \boldsymbol{\xi}^{tr,r}_i(t)  \boldsymbol{\xi}^{tr,r}_{j}(t^{\prime}) \rangle &=   
 \mathbb{I}_3 \, \delta_{ij} \, \delta(t-t^{\prime}),\label{variance2}
 \end{align}
where $ \mathbb{I}_3$ is the unit matrix in three dimensions.

We implemented the aforementioned equations of motion (Eqs. \ref{langevin} and  \ref{roteq}) using an Euler-Maruyama  integration scheme.\cite{Higham2001} A maximum time step of  $dt=2\times 10^{-5} \sigma^2 /D_{tr}$ was used for the numeric integration of the equations of motion. The number of particles in our simulations was approximately $N=2500$, and we have verified that our results are robust upon doubling the number of particles. 

Lengths are given in units of $\sigma$, time in units of $\tau=\sigma^2 /D_{tr}$, and energy in units of $1/\beta_s$. We also denote $T={1}/{\beta_s\epsilon}$ as the dimensionless temperature of our system. This notation is adopted as it facilitates direct comparison to a passive LJ system. 

\subsection{Pressure} \label{Analysismethods}

In order to measure the pressure of our active system we use the results of Winkler \textit{et al}.\cite{winkler2015virial} Specifically, the pressure $P$ of a system of self-propelled and isotropic particles in a periodic box is calculated using
\begin{equation}
 P=P_{id}+P_{vir}+P_{swim}.
\end{equation}
In this expression, the ideal gas pressure $P_{id}$ is  given by
\begin{equation}
 P_{id}=\rho/\beta_s,\label{eq:idealpressure}
\end{equation}
 with $\rho$  the number density. Additionally, $P_{vir}$ is the standard virial pressure given by
\begin{equation}\label{eq:virialpressure}
 P_{vir}=-\frac{1}{3V}\left\langle \sum_{i=1}^{N-1} \sum_{j= i+1}^N\frac{\partial U(r_{ij})}{\partial \boldsymbol{r}_i}  \cdot(\boldsymbol{r}_i -\boldsymbol{r}_j) \right\rangle ,
\end{equation}
 where $V$ is the volume of the system. Finally, $P_{swim}$ is the ``swim pressure'', i.e. the direct contribution of the self-propulsive forces to  the pressure, and is given by
\begin{align}\label{eq:swimpressure}
 P_{swim}=&\frac{\rho \eta \upsilon_0^2}{6 D_r} \nonumber
 \\
 &-\frac{\eta\upsilon_0}{6VD_r}\left\langle \sum_{i=1}^{N-1} \sum_{j= i+1}^N\frac{\partial U(r_{ij})}{\partial \boldsymbol{r}_i}  \cdot  (  \boldsymbol{u}_i - \boldsymbol{u}_j ) \right\rangle  .
\end{align}
Note that the brackets in Eqs. \ref{eq:virialpressure} and \ref{eq:swimpressure} denote a time average over the steady state. The steady state of the system was identified following Ref. \onlinecite{prymidis2015self}. 

\section{Results}

%
%
%
%

\subsection{Equations of state}
\label{equationsofstate}

Recent theoretical work has established the existence of an equation of state for isotropic, self-propelled particles, such as our model.\cite{takatori2014swim, winkler2015virial, falasco2015mesoscopic, solon2015pressure} In this section we calculate equations of state for an active LJ system in a periodic cubic box. Our goal is to examine the behaviour of the equation of state as the active system transitions from a homogeneous state to vapour-liquid phase coexistence, and compare it with the behaviour of a passive LJ system. 

\begin{figure}
\includegraphics[scale=0.45]{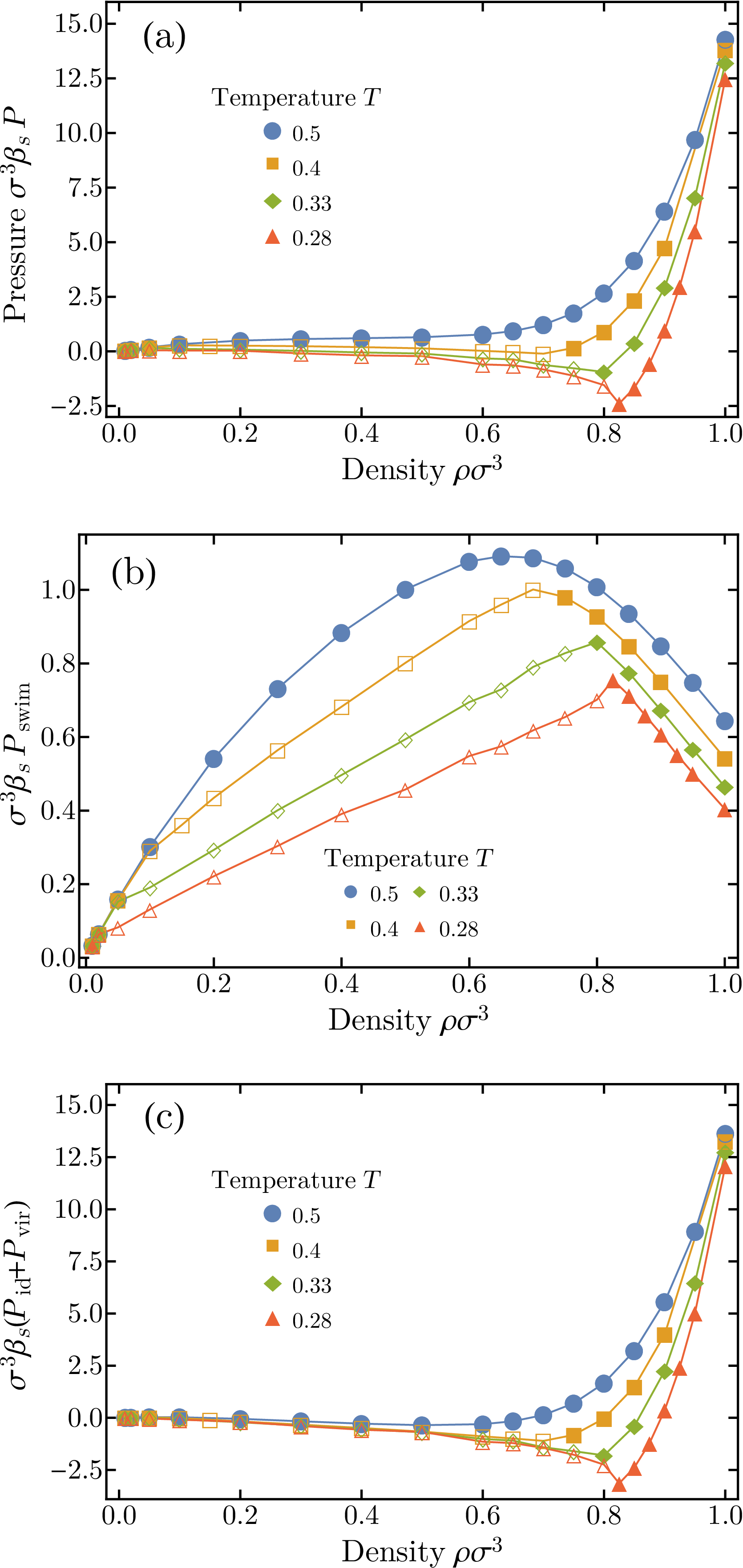}
\caption{\label{fig:eos} Equations of state for a system with propulsion speed $\upsilon_0 \tau/\sigma=20$ and rotational diffusion coefficient $D_r\tau=20$. (a) shows the total pressure of the system as a function of the density, (b) shows  the swim pressure contribution (Eq. \ref{eq:swimpressure}) and (c) shows the sum of the ideal and the virial contribution (Eqs.  \ref{eq:idealpressure} and \ref{eq:virialpressure} respectively). Full symbols correspond to state points where the system is in a homogeneous state while open symbols denote vapour-liquid phase coexistence.  Full lines are simply guides to the eye.  }
\end{figure}

In Fig. \ref{fig:eos}(a) we show characteristic  equations of state for the system. As in the passive system, lowering the temperature causes the equation of state to become non-monotonic, a behaviour associated with phase separation into a gas and a liquid. From these equations of state, we observe no qualitative differences from the passive LJ system. 

In order to examine the equations of state in more detail, we study the different contributions to the pressure. In Fig. \ref{fig:eos}(b) we plot the swim pressure as a function of the density for different temperatures. We find that for high temperatures, where no coexistence takes place, the swim pressure has a roughly parabolic shape. However,  once phase separation occurs in the system, the swim pressure grows linearly with the density in the phase coexistence regime.
This linear growth in the coexistence region is present  for all other parameter space points that we have examined. We find that the swim pressure of both the gas and the liquid phase stays fixed throughout the coexistence region, hence this linear growth of the total swim pressure arises due to the lever rule.

Subsequently, in Fig. \ref{fig:eos}(c) we show the contribution coming from the ideal and the passive virial part of the pressure. Note that these two contributions alone cannot account for the observed phase behaviour, as the high temperature curve (colored blue) is non-monotonic  even though the system is in a fluid state for all densities shown.

%
%
%
%

\subsection{Vapour-Liquid Coexistence}
\label{gasliquidcoexistence}

\begin{figure}
\includegraphics[scale=0.45]{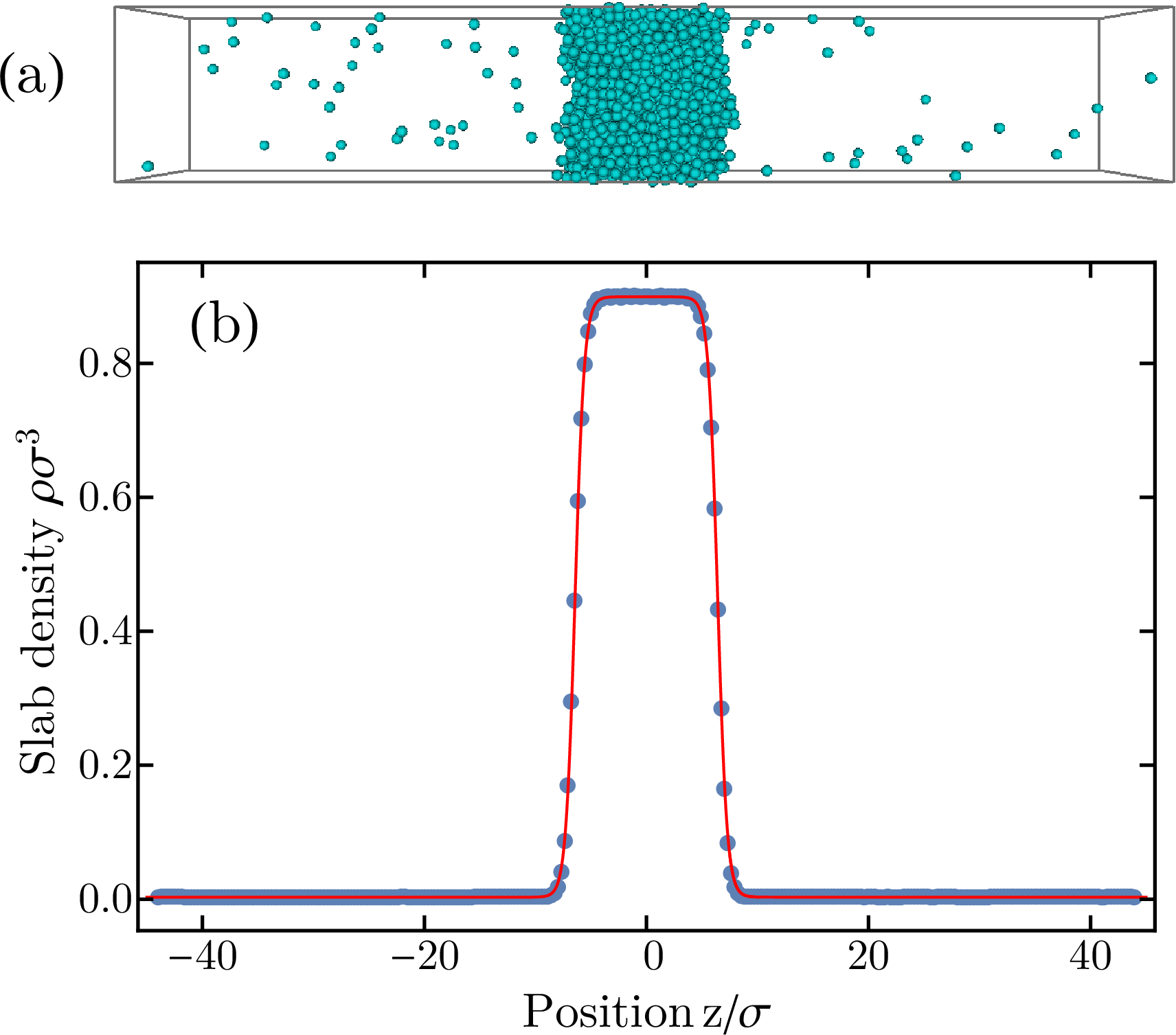}
\caption{\label{fig:screens} (a) Snapshot of a long box simulation. (b) Average number density $\rho \sigma^3 $ as a function of position. Data points are the time averages obtained from simulations while the red curve is a fit of Eq. \ref{eq:fitdensity}. }
\end{figure}

In this section we map out the phase diagram for the LJ fluid. 
To this end, we conducted simulations in a long simulation box with dimensions $L_z=6L_x=6L_y$, containing a liquid slab coexisting with vapour, as shown in Figure \ref{fig:screens}(a). The overall number density of the system was fixed at $\rho\sigma^3=0.1333$.  We then measured the density profile along the long axis by dividing the box into slabs of width $\simeq 0.3\sigma$ along the $z$ direction and taking the time average of the number of particles in a given slab. Subsequently, we calculated the local number densities of the vapour phase $\rho_v$ and the liquid phase $\rho_l$  by fitting the density profile  $\rho(z)$ around each interface to the function
\begin{equation}\label{eq:fitdensity}
 \rho(z)=\frac{1}{2}(\rho_l+\rho_v)-\frac{1}{2}(\rho_l-\rho_v)\tanh\left[  \frac{2(z-z_0)}{w}  \right],
\end{equation}
where $z_0$ and $w$ are the location and width of the vapour-liquid interface, and are also determined from the fit. 
Fig. \ref{fig:screens}(b) shows an example of a measured density profile as well as the fitted Eq. \ref{eq:fitdensity}. We find that the hyperbolic tangent provides an excellent fit to the interface and that we can accurately determine the local densities of the vapour and the liquid phase.

\begin{figure}
 \includegraphics[scale=0.32]{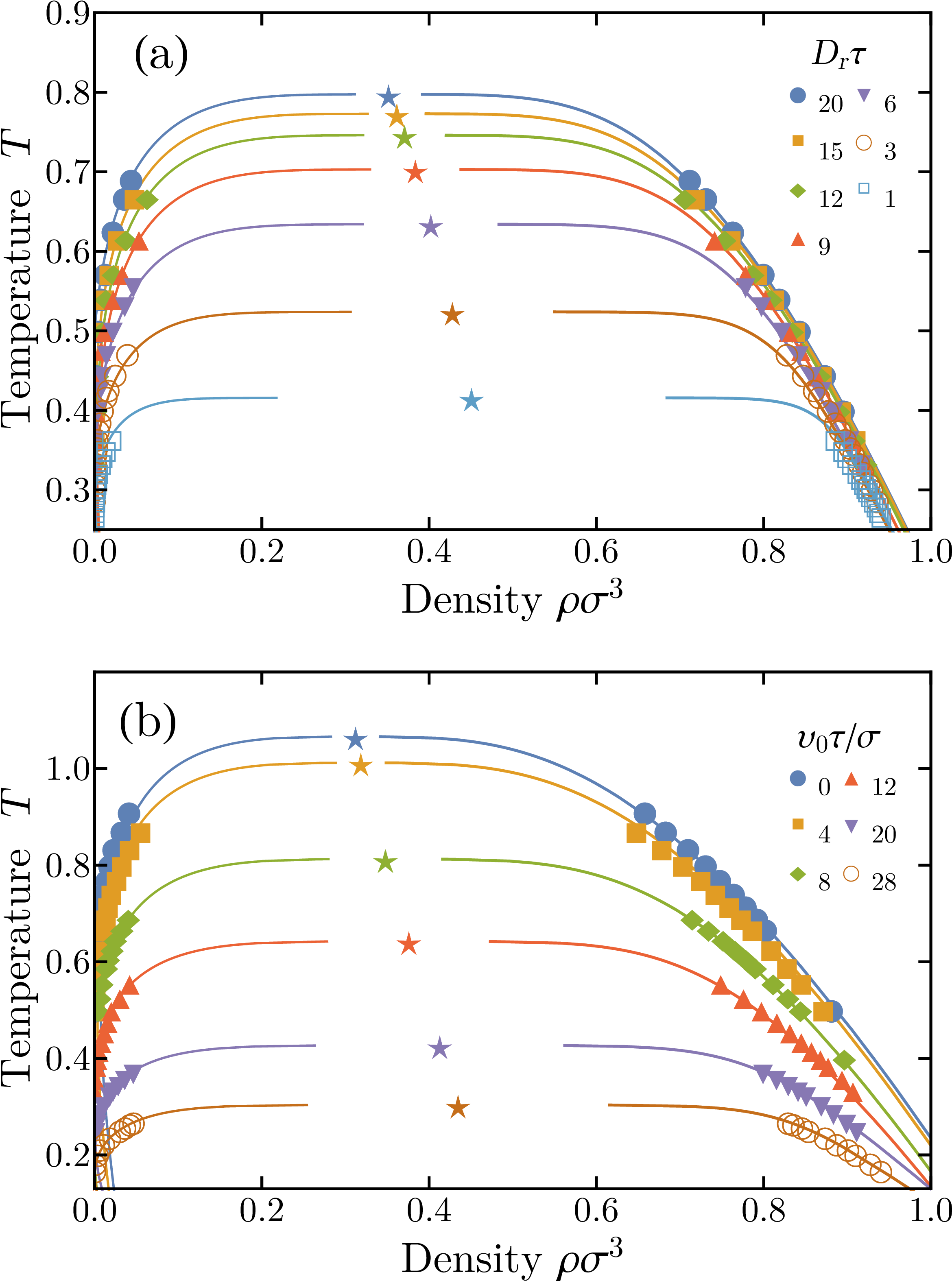}
\caption{\label{fig:binodals}Binodal lines of the system for a system of (a) constant propulsion speed $\upsilon_0 \tau/\sigma=8$ and varying rotational diffusion coefficient  and (b) constant rotational diffusion coefficient $D_r \tau=20$  and varying propulsion speed. Simulation results are denoted by points. Full lines  are fits, obtained by using the exponential fits for the parameters $T_c, \beta, A, \rho_c, \alpha$ and $B$ (Tables \ref{tab:coef} and \ref{tab:spe}) on Eqs. \ref{eq:deltarho} and \ref{eq:plusrho}. Stars denote the calculated critical points. }
\end{figure}

We systematically obtained the coexisting densities for a wide range of parameters following two different paths that drive the system out of equilibrium. First, we varied the rotational diffusion coefficient while keeping the self-propulsion fixed at a non-zero value. Second,  we varied the propulsion speed while keeping the rotational diffusion rate of the particles fixed. The measured coexisting densities are summarized in Figure \ref{fig:binodals}. Clearly, both routes produce a series of phase diagrams that are highly consistent with a simple passive attractive fluid, such as a LJ fluid. 
 Note that these two paths are not equivalent as the $D_r\rightarrow \infty$ limit does not coincide with the  $\upsilon_0\rightarrow 0$ limit: the first one corresponds  to a passive system  with a higher effective temperature than the second one, which corresponds to the equilibrium LJ system with temperature $T=1/\beta_s\epsilon$. 

 In the following subsections, we compare the obtained phase diagrams more closely to the equilibrium case by exploring the temperature dependence of $\Delta \rho = \rho_l-\rho_v$, and examining whether the law of rectilinear diameters and law of corresponding states still hold.
 
%
%
%
%

\subsubsection{Temperature dependence of $\Delta \rho$}
\label{orderparameter}

The order parameter that governs the vapour-liquid phase transition in equilibrium is the difference between the two coexisting densities $\Delta\rho=\rho_l-\rho_v$. In equilibrium, $\Delta\rho$ follows a power law given by
\begin{equation}\label{eq:deltarho}
  \sigma^3 \Delta\rho = A \left(T_c-T \right)^{\beta},  
\end{equation}
where $T_c$ is the critical temperature, $\beta$ is the (critical) exponent and $A$ is a proportionality constant. Here we examine whether the scaling of $\Delta \rho$ with temperature follows the same behaviour for our active system, and treat $T_c$, $A$, and $\beta$ as free fitting parameters.

\begin{figure}
 \includegraphics[scale=0.25]{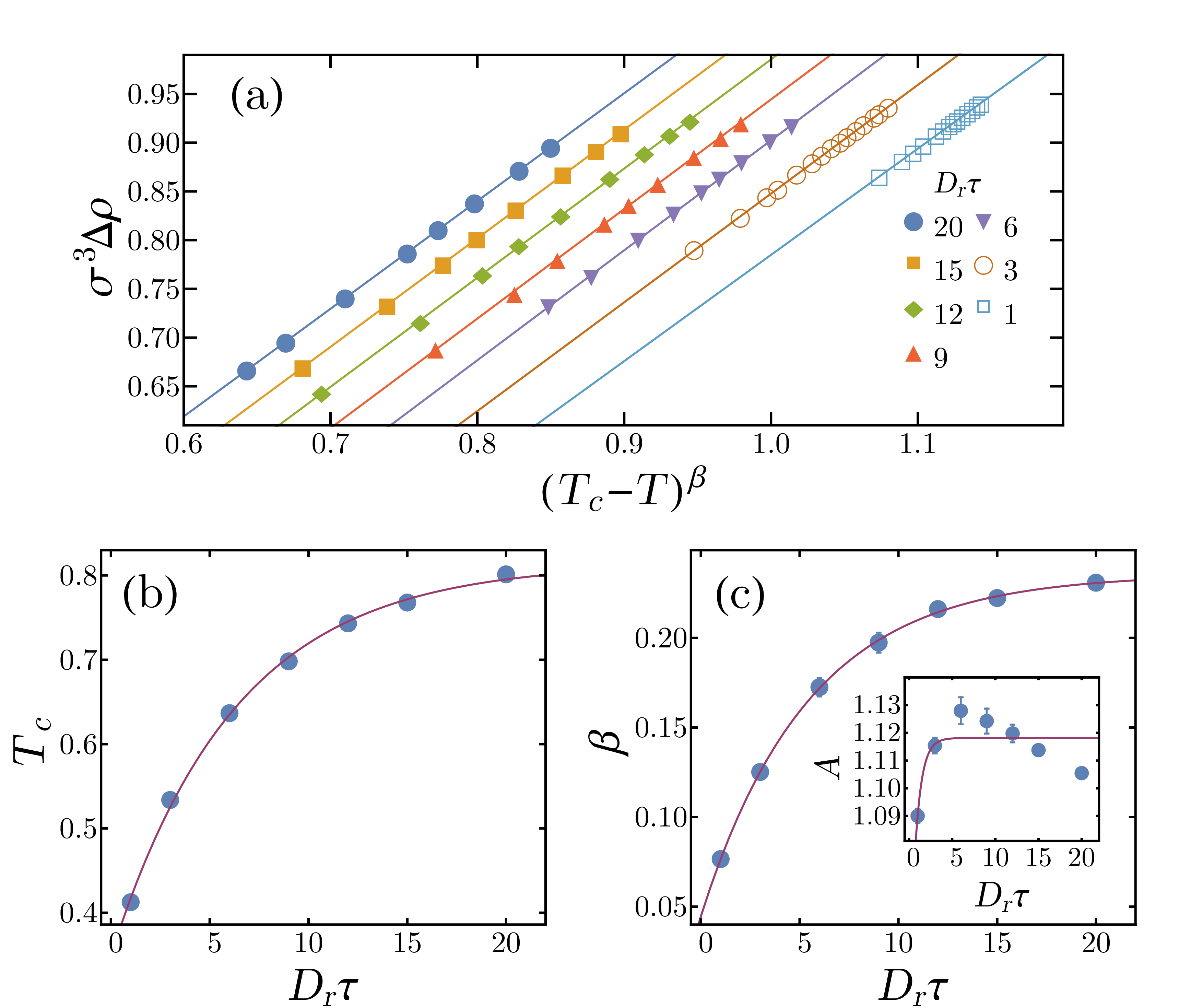}
\caption{\label{fig:deltarhorot} (a) $\Delta \rho$ as a function of the scaled temperature for systems of different rotational diffusion rates and constant propulsion speed $\upsilon_0 \tau/\sigma=8$. Data points correspond to simulation results while the lines denote the fits (Eq. \ref{eq:deltarho}). Results for different rotational diffusion rates are offset for clarity. (b) Critical temperature $T_c$ as a function of the rotational diffusion rate. The continuous line shows the fit from Table \ref{tab:coef} . (c) Critical exponent $\beta$ and constant $A$ (inset) as a function of the rotational diffusion rate. The continuous line shows the fit from Table \ref{tab:coef}.}
\end{figure}

\begin{figure}
\includegraphics[scale=0.25]{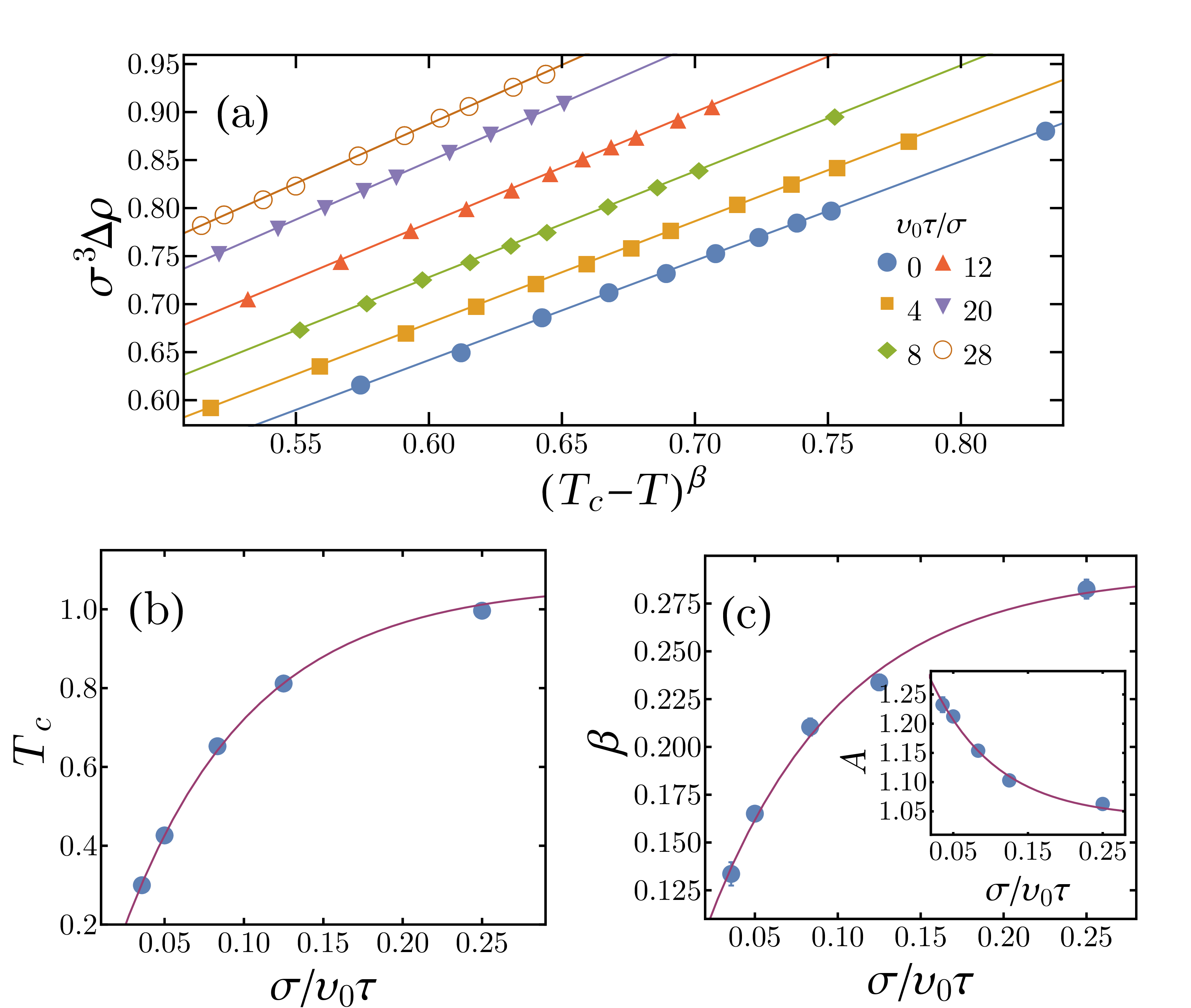}
\caption{ \label{fig:deltarhospe} (a) $\Delta \rho$ as a function of the scaled temperature for systems of different propulsion speeds and constant rotational diffusion $D_r\tau=20$. Data points correspond to simulation results while the lines denote the fits (Eq. \ref{eq:deltarho}). Results for different propulsion speeds are offset for clarity. (b) Critical temperature $T_c$ as a function of the self-propulsion speed. The continuous line shows the fit from Table \ref{tab:spe}. (c) Critical exponent $\beta$ and constant $A$ (inset) as a function of the self-propulsion speed. The continuous line shows the fit from Table \ref{tab:spe}.}
\end{figure}

In Figs. \ref{fig:deltarhorot}(a) and  \ref{fig:deltarhospe}(a)  we show the order parameter $\Delta\rho$, as a function of the  scaled temperature (Eq. \ref{eq:deltarho}) for different values of rotational diffusion rate and self-propulsion speed, respectively. Interestingly, the simulation data  fall on straight lines, indicating that Eq. \ref{eq:deltarho}  accurately describes the active system in the examined parameter space. 

Next, we  examine the scaling of the fitted critical temperature $T_c$, the exponent $\beta$ and the constant $A$ as the system is driven away from equilibrium.  The results are plotted in Figs. \ref{fig:deltarhorot}(b-c) and   \ref{fig:deltarhospe}(b-c). Error bars are the standard errors from fitting and in the majority of cases they are smaller than the plotted markers. As expected, the critical temperature decreases with decreasing rotational diffusion coefficient/increasing  propulsion speed as stronger attraction is needed to bring together swimmers with larger persistence lengths. We also find that the  exponent $\beta$ decreases as our particles become more active. The parameter $A$ stays quasi-constant as a function of the rotational diffusion (Fig. \ref{fig:deltarhorot}(c) inset), but clearly increases with increasing propulsion speed (Fig. \ref{fig:deltarhospe}(c) inset).

For the systems where the rotational diffusion coefficient is varied, the scaling of $T_c$ and $\beta$ is well captured by simple exponential functions, for instance
\begin{align}
 T_c(D_r )&=a_1+a_2e^{-a_3 {D_r\tau}} \label{eq:tcscaling},
\end{align}
is an excellent fit for the critical temperature, where the values of the dimensionless parameters $a_1, a_2$ and $a_3$ can be found in Table \ref{tab:coef}.  We also fit the parameter $A$ with an exponential function, even though its variation is minimal and the fit is clearly not optimal.

For the systems where the propulsion speed is varied, we similarly find that the scaling
\begin{align}
 T_c(\upsilon_0)&=b_1+b_2e^{-b_3 \sigma/(\upsilon_0 \tau) } \label{eq:tcscaling2},
\end{align}
describes our data fairly well. The same holds for the exponent $\beta$ and the constant $A$. The numerical coefficients can be found in Table \ref{tab:spe}. Note that the difference between Eqs. \ref{eq:tcscaling2} and \ref{eq:tcscaling} is simply the replacement of $D_r \tau$ with $\sigma/(\upsilon_0 \tau)$.  As the P\'eclet number is simply the ratio of these two, it might be tempting to ask whether the phase behaviour can be completely described by the P\'eclet number $\mathrm{Pe}=D_r \sigma / \upsilon_0$.  However, this turns out not to be the case as these two separate paths out of equilibrium cannot be collapsed via the P\'eclet number.

\begin{table}
\caption{\label{tab:coef} Fitting parameters of the function $a_1+a_2e^{-a_3 {D_r\tau}}$ to the parameters of Eqs. \ref{eq:deltarho} and \ref{eq:plusrho} for systems of different rotational diffusion rates and constant propulsion speed $\upsilon_0 \tau/\sigma=8$.}
\begin{tabular}{ |c || c |  c |  c |  c |  c | c |}
 \hline  & $T_c$ & $\beta$ & $A$ & $\rho_c\sigma^3$  & $\alpha$ & $B$  \\
\hhline{|=#=|=|=|=|=|=|}
$a_1$ & 0.818 & 0.237 & 1.118 & 0.339 & 1.04 & 0.252 \\
\hline $a_2$ & -0.47 & -0.194 & -0.081 & 0.126 & 1.122 & 1.336 \\
\hline $a_3$ & 0.156 & 0.18 & 1.004 & 0.114 & 0.229 & 0.79 \\
\hline
\end{tabular}
\end{table}

\begin{table}
\caption{\label{tab:spe} Fitting parameters of the function $b_1+b_2e^{-b_3  \sigma/(\upsilon_0 \tau) }$ to the parameters of Eqs. \ref{eq:deltarho} and \ref{eq:plusrho} for systems of different propulsion speeds and constant rotational diffusion $D_r\tau=20$.}
\begin{tabular}{ |c || c |  c |  c |  c |  c | c |}
 \hline  & $T_c$ & $\beta$ & $A$ &  $\rho_c\sigma^3$ & $\alpha$ & $B$  \\
\hhline{|=#=|=|=|=|=|=|}
$b_1$ & 1.066 & 0.291 & 1.038 & 0.312 &0.971  &0.234  \\
\hline $b_2$ & -0.184  &-0.241  &0.297  &0.201 &2.132 &8.082 \\
\hline $b_3$ & 12.33&12.382 &11.384 &13.795 &33.406 & 69.521 \\
\hline
\end{tabular}
\end{table}

%
%
%
%
\subsubsection{Law of rectilinear diameters}
\label{lawofrectilinear}

\begin{figure}
 \includegraphics[scale=0.25]{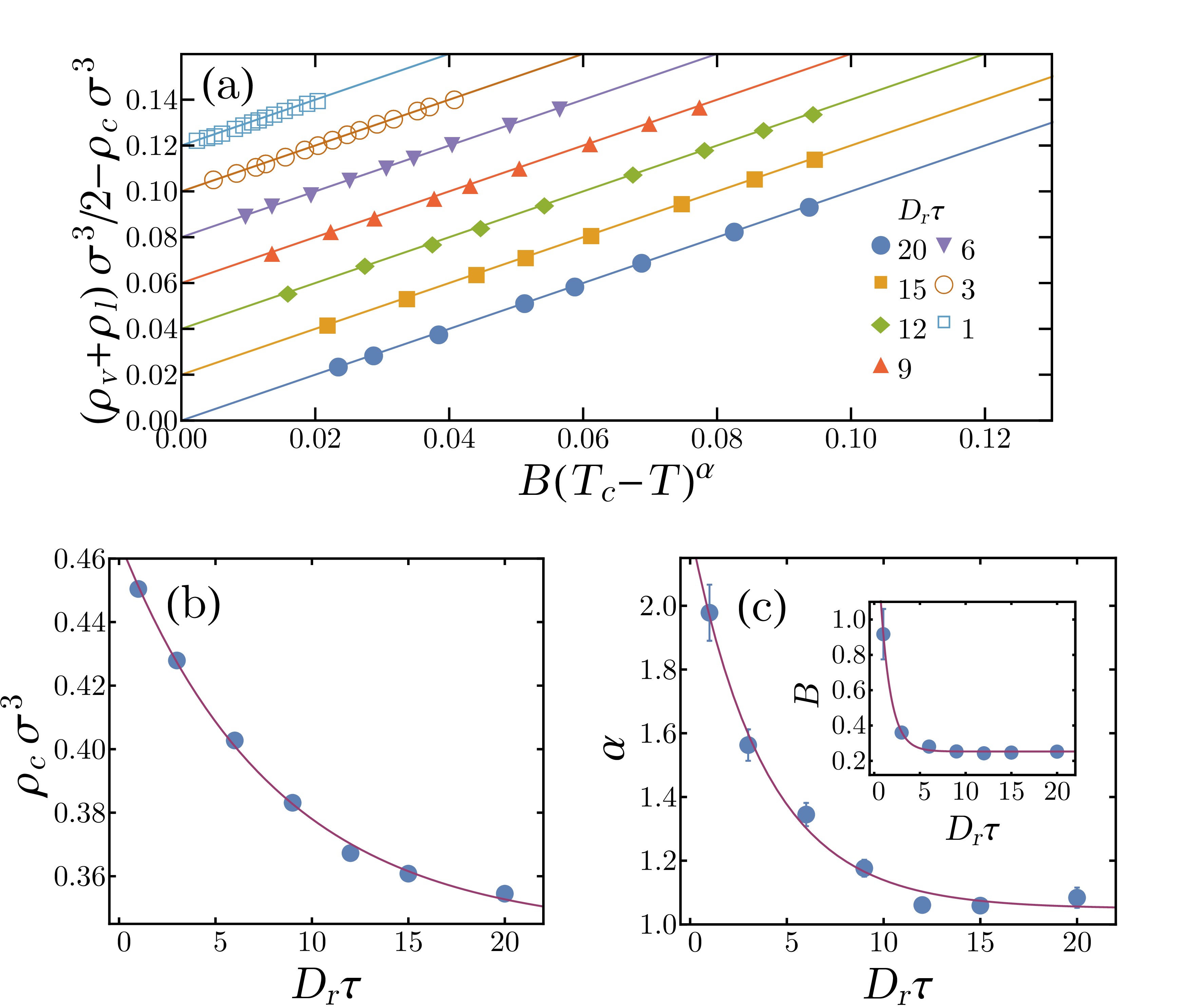}
\caption{\label{fig:rhoplusro} Distance of the average of the coexisting densities from the critical density $\frac{1}{2}(\rho_v + \rho_l)-\rho_c$ as a function of the scaled temperature for different rates of rotational diffusion and constant self-propulsion speed $\upsilon_0 \tau/\sigma=8$.  Data points correspond to simulation results while the lines denote the fits of Eq. \ref{eq:plusrho}. Results for different rotational diffusion rates are offset for clarity. (b) Critical density $\rho_c$  as a function of the rotational diffusion coefficient $D_r$. The continuous line shows the fit from Table \ref{tab:coef}. (c)  Critical exponent $\alpha$ and constant $B$ (inset)  as a function of the rotational diffusion coefficient $D_r$. The continuous line shows the fit from Table \ref{tab:coef}.}
\end{figure}

\begin{figure}
\includegraphics[scale=0.25]{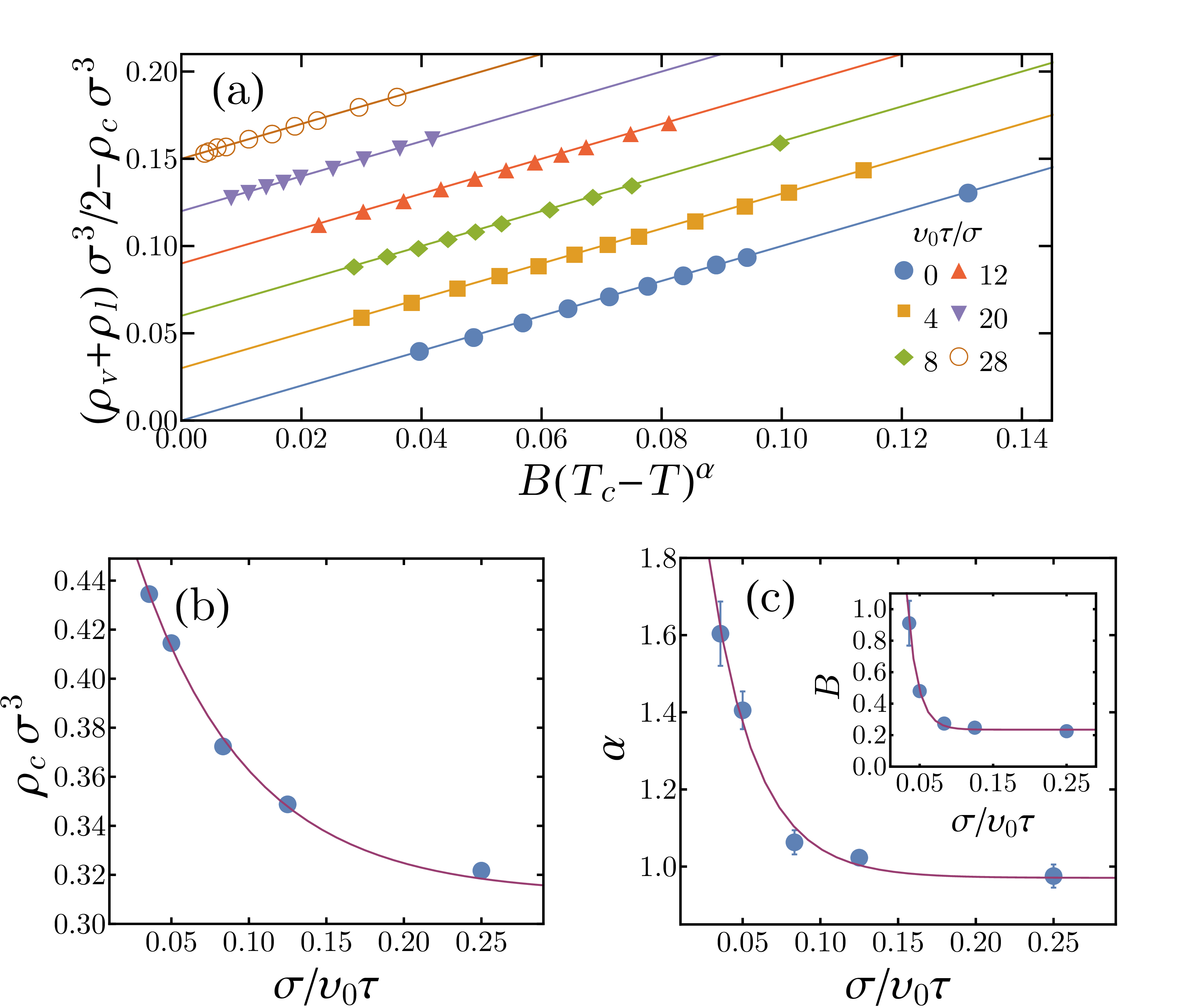}
\caption{\label{fig:rhopluspe} Distance of the average of the coexisting densities from the critical density $\frac{1}{2}(\rho_v + \rho_l)-\rho_c$ as a function of the scaled temperature for different propulsion speeds and constant rotational diffusion coefficient $D_r\tau=20$.  Data points correspond to simulation results while the lines denote the fits of Eq. \ref{eq:plusrho}. Results for different propulsion speeds are offset for clarity. (b) Critical density $\rho_c$  as a function of the self-propulsion speed. The continuous line shows the fit from Table \ref{tab:spe}.(c) Critical exponent $\alpha$ and constant $B$ (inset)  as a function of the self-propulsion speed. The continuous line shows the fit from Table \ref{tab:spe}.}
\end{figure}

Next, we investigate whether the law of rectilinear diameters holds for our system. Specifically, we  study the properties of the sum of the coexisting densities which in equilibrium \cite{guggenheim1945principle, vrabec2006comprehensive} typically scales as
\begin{equation}
 \label{eq:plusrho}
  \frac{1}{2}\left(\rho_v+\rho_l\right)\sigma^3=B\left(T_c-T\right)^{\alpha}+\rho_c\sigma^3.
\end{equation}
with $B$ a proportionality constant, $\alpha$ the exponent and $\rho_c$ the density at the critical point. According to the law of rectilinear diameters,  the exponent $\alpha=1$.  Note that in Eq. \ref{eq:plusrho} we have omitted corrections that are needed in order to capture the behaviour near the critical point, as we are unable to study this regime in the present work. Using $T_c$ as calculated in the previous section, we determine the proportionality constant $B$, the exponent $\alpha$ and the critical density $\rho_c$ by fitting the coexisting densities to this expression. In Figs. \ref{fig:rhoplusro}(a) and \ref{fig:rhopluspe}(a) we show that Eq. \ref{eq:plusrho} can indeed accurately reproduce the behaviour of our out-of-equilibrium system as we vary the rotational diffusion rate and self-propulsion speed respectively.

Next, in Figs. \ref{fig:rhoplusro}(b-c) and \ref{fig:rhopluspe}(b-c) we plot the fitted parameters critical density $\rho_c$, exponent $\alpha$ and constant $B$. In Figs.  \ref{fig:rhoplusro}(b) and \ref{fig:rhopluspe}(b) one can see that the critical density increases with decreasing rotational diffusion coefficient/increasing propulsion speed, indicating that higher densities are necessary in order to have coexistence when particles swim faster. Interestingly, we also find that the exponent $\alpha$ deviates substantially from unity as we drive the system away from equilibrium. Thus, sufficiently far from equilibrium the law of rectilinear diameters is clearly violated. Lastly, the parameter $B$ also increases with  decreasing rotational diffusion coefficient/increasing propulsion speed.

In addition, the three parameters $\rho_c$, $\alpha$ and $B$ can be fitted  again with a simple exponential of the form of Eq. \ref{eq:tcscaling} or \ref{eq:tcscaling2}, depending on whether the rotational diffusion or the propulsion speed is varied. The measured fits can be found in Tables \ref{tab:coef} and \ref{tab:spe}. The fact that all fitting parameters $T_c, \beta,  \rho_c, \alpha$ and $B$ scale in a similar fashion in the active system is remarkable, and may suggest that a simple, comprehensive description of the phase transition is indeed possible for our model.

%
%
%
%
\subsubsection{Binodal lines and law of corresponding states}
\label{binodals}

Finally, one can now combine Eqs. \ref{eq:deltarho} and \ref{eq:plusrho} in order to express the coexisting densities $\rho_v$ and $\rho_l$ as a function of the parameters $T_c, \beta, A, \rho_c, \alpha$ and $B$. In Fig. \ref{fig:binodals} we compare the binodals of the system from the directly measured coexisting densities to the fits for the aforementioned parameters (Tables \ref{tab:coef} and \ref{tab:spe}).  We find that the agreement between measurements and  fits is excellent. 
We note that in Ref. \onlinecite{prymidis2015self} a percolating network state separated the fluid from the vapour-liquid coexistence region when the system was sufficiently far from equilibrium. Consequently, this extra state may well result in a metastable critical point for our system. However, we have performed simulations at all the predicted critical temperatures and observed no signatures of a percolating state within the predicted coexistence regions.

\begin{figure}
\includegraphics[scale=0.35]{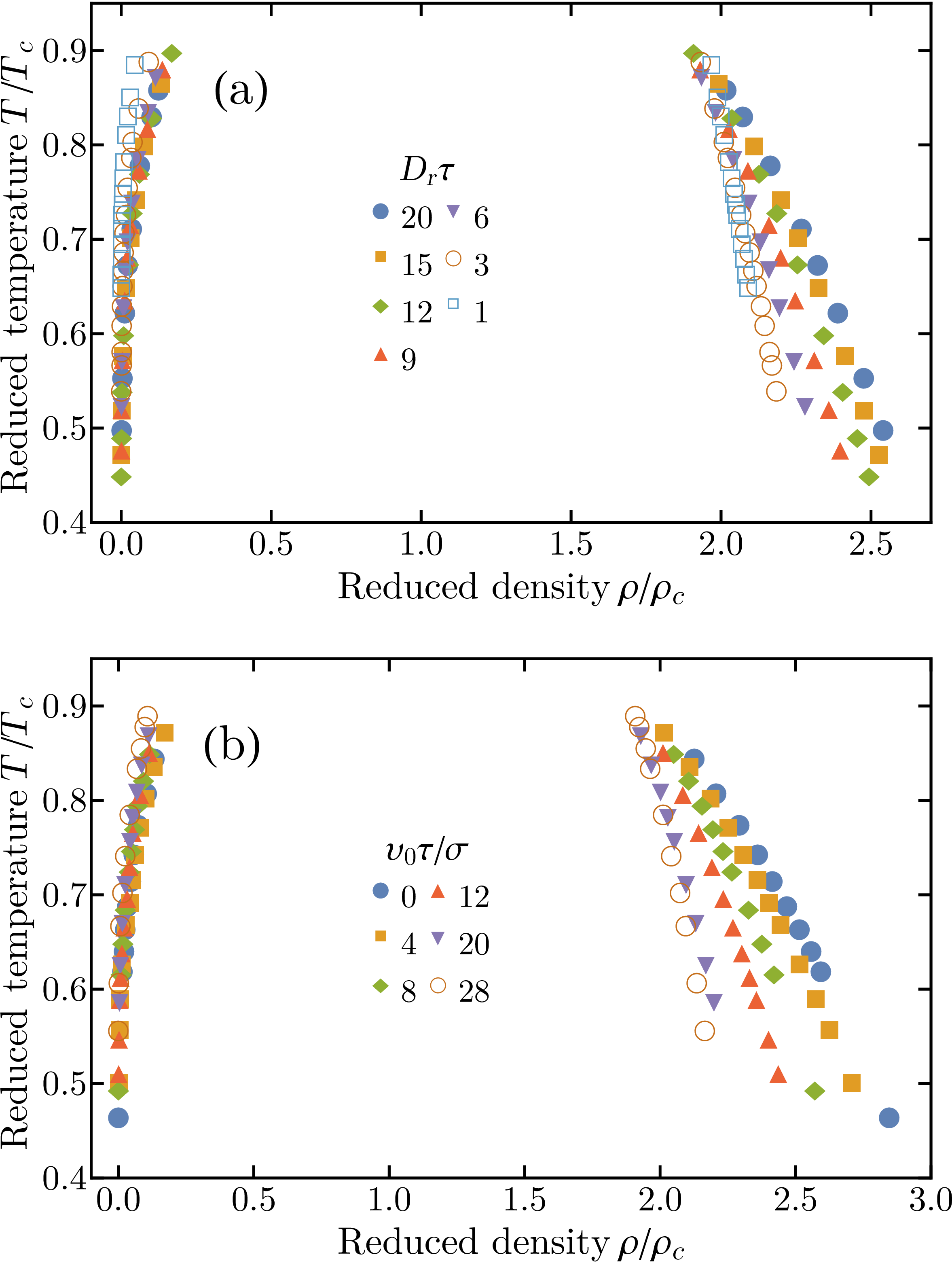}
\caption{\label{fig:corre} Reduced binodal lines of the active LJ system (a) for systems of different rotational diffusion rates and constant propulsion speed $\upsilon_0 \tau/\sigma=8$ and (b) for systems of different  propulsion speeds and constant rotational diffusion coefficient $D_r\tau=20$. Points correspond to directly observed coexisting densities, scaled by the fitted critical temperatures and densities. }
\end{figure}

Last but not least, we checked whether our system obeys  a simple law of corresponding states. That is, whether the binodal lines fall on top of each other if one scales the temperature and the density with the corresponding quantities at the critical point. Such a collapse of the binodals can be made, for example, for various real substances,\cite{guggenheim1945principle} or for different  cutoff radii of the equilibrium LJ fluid.\cite{dunikov2001corresponding} However, as shown in Figure \ref{fig:corre}, the active LJ fluid obeys no such law of corresponding states for different values of  rotational diffusion and self-propulsion. Naturally, the fact that the active LJ fluid does not obey this simplified law of corresponding states does not prove that it does not obey a more general Noro-Frenkel law of corresponding states, which compares the thermodynamic properties at the same reduced density and second virial coefficient. However, a more general test is out of the scope of the present work. 

%
%
%
%
\section{Conclusions}
\label{Discussion}

We studied a system of self-propelled spheres that interact via the Lennard-Jones potential using Brownian Dynamics simulations. We  calculated equations of state for different temperatures and verified that, as the system transitions from a homogeneous to a phase separated state with decreasing temperature, the pressure curve as a function of density shows the expected transition from  monotonic to non-monotonic. Moreover, we observed a linear growth of the swim pressure in the coexistence region.

Subsequently, we studied the phase coexistence regime using long box simulations. We showed that the scaling of the coexisting densities with temperature follows classic power laws.  Though it might be tempting to identify the exponents of these power laws with the critical exponents of the system, we are currently unable to access the region close to the critical point due to the small system sizes considered here. Hence we cannot be certain that the calculated power laws still hold in this region.  Nonetheless, the power laws we present describe the binodal envelope extremely well and should provide guidance for future studies of the system. 

Noticeably, we also showed that all the various parameters of the power laws vary with the propulsion speed or the rotational diffusion rate in a similar fashion, namely their scaling is well captured by simple exponential functions. These parameters include the critical temperature and density as well as the exponents of the power laws. Thus, a unified description of the binodal lines for both the passive and the active Lennard-Jones system may be within reach.

\begin{acknowledgments}
V.P. and L.F. acknowledge funding from the Dutch Sector Plan Physics and Chemistry, and L. F. acknowledges financial support from the Netherlands Organization for Scientific Research (NWO-VENI grant No. 680.47.432). S.P. and M.D. acknowledge the funding from the Industrial Partnership Programme ``Computational  Sciences  for  Energy  Research"  (Grant  No.14CSER020) of the Foundation for Fundamental Research on Matter (FOM), which is part of the Netherlands Organization for Scientific Research (NWO). This research programme is co-financed by Shell Global Solutions International B.V. The authors would like to thank Frank Smallenburg for careful reading of the manuscript and Robert Evans for useful discussions. 
\end{acknowledgments}

\bibliography{Paper.bib}

\begin{thebibliography}{43}%
\makeatletter
\providecommand \@ifxundefined [1]{%
 \@ifx{#1\undefined}
}%
\providecommand \@ifnum [1]{%
 \ifnum #1\expandafter \@firstoftwo
 \else \expandafter \@secondoftwo
 \fi
}%
\providecommand \@ifx [1]{%
 \ifx #1\expandafter \@firstoftwo
 \else \expandafter \@secondoftwo
 \fi
}%
\providecommand \natexlab [1]{#1}%
\providecommand \enquote  [1]{``#1''}%
\providecommand \bibnamefont  [1]{#1}%
\providecommand \bibfnamefont [1]{#1}%
\providecommand \citenamefont [1]{#1}%
\providecommand \href@noop [0]{\@secondoftwo}%
\providecommand \href [0]{\begingroup \@sanitize@url \@href}%
\providecommand \@href[1]{\@@startlink{#1}\@@href}%
\providecommand \@@href[1]{\endgroup#1\@@endlink}%
\providecommand \@sanitize@url [0]{\catcode `\\12\catcode `\$12\catcode
  `\&12\catcode `\#12\catcode `\^12\catcode `\_12\catcode `\%12\relax}%
\providecommand \@@startlink[1]{}%
\providecommand \@@endlink[0]{}%
\providecommand \url  [0]{\begingroup\@sanitize@url \@url }%
\providecommand \@url [1]{\endgroup\@href {#1}{\urlprefix }}%
\providecommand \urlprefix  [0]{URL }%
\providecommand \Eprint [0]{\href }%
\providecommand \doibase [0]{http://dx.doi.org/}%
\providecommand \selectlanguage [0]{\@gobble}%
\providecommand \bibinfo  [0]{\@secondoftwo}%
\providecommand \bibfield  [0]{\@secondoftwo}%
\providecommand \translation [1]{[#1]}%
\providecommand \BibitemOpen [0]{}%
\providecommand \bibitemStop [0]{}%
\providecommand \bibitemNoStop [0]{.\EOS\space}%
\providecommand \EOS [0]{\spacefactor3000\relax}%
\providecommand \BibitemShut  [1]{\csname bibitem#1\endcsname}%
\let\auto@bib@innerbib\@empty
\bibitem [{\citenamefont {Ramaswamy}(2010)}]{ramaswamy2010mechanics}%
  \BibitemOpen
  \bibfield  {author} {\bibinfo {author} {\bibfnamefont {S.}~\bibnamefont
  {Ramaswamy}},\ }\href@noop {} {\bibfield  {journal} {\bibinfo  {journal}
  {Annu. Rev. Condens. Matter Phys.}\ }\textbf {\bibinfo {volume} {1}},\
  \bibinfo {pages} {323} (\bibinfo {year} {2010})}\BibitemShut {NoStop}%
\bibitem [{\citenamefont {Romanczuk}\ \emph {et~al.}(2012)\citenamefont
  {Romanczuk}, \citenamefont {B{\"a}r}, \citenamefont {Ebeling}, \citenamefont
  {Lindner},\ and\ \citenamefont {Schimansky-Geier}}]{romanczuk2012active}%
  \BibitemOpen
  \bibfield  {author} {\bibinfo {author} {\bibfnamefont {P.}~\bibnamefont
  {Romanczuk}}, \bibinfo {author} {\bibfnamefont {M.}~\bibnamefont {B{\"a}r}},
  \bibinfo {author} {\bibfnamefont {W.}~\bibnamefont {Ebeling}}, \bibinfo
  {author} {\bibfnamefont {B.}~\bibnamefont {Lindner}}, \ and\ \bibinfo
  {author} {\bibfnamefont {L.}~\bibnamefont {Schimansky-Geier}},\ }\href@noop
  {} {\bibfield  {journal} {\bibinfo  {journal} {Eur. Phys. J. Special Topics}\
  }\textbf {\bibinfo {volume} {202}},\ \bibinfo {pages} {1} (\bibinfo {year}
  {2012})}\BibitemShut {NoStop}%
\bibitem [{\citenamefont {Marchetti}\ \emph {et~al.}(2013)\citenamefont
  {Marchetti}, \citenamefont {Joanny}, \citenamefont {Ramaswamy}, \citenamefont
  {Liverpool}, \citenamefont {Prost}, \citenamefont {Rao},\ and\ \citenamefont
  {Simha}}]{marchetti2013hydrodynamics}%
  \BibitemOpen
  \bibfield  {author} {\bibinfo {author} {\bibfnamefont {M.}~\bibnamefont
  {Marchetti}}, \bibinfo {author} {\bibfnamefont {J.}~\bibnamefont {Joanny}},
  \bibinfo {author} {\bibfnamefont {S.}~\bibnamefont {Ramaswamy}}, \bibinfo
  {author} {\bibfnamefont {T.}~\bibnamefont {Liverpool}}, \bibinfo {author}
  {\bibfnamefont {J.}~\bibnamefont {Prost}}, \bibinfo {author} {\bibfnamefont
  {M.}~\bibnamefont {Rao}}, \ and\ \bibinfo {author} {\bibfnamefont {R.~A.}\
  \bibnamefont {Simha}},\ }\href@noop {} {\bibfield  {journal} {\bibinfo
  {journal} {Rev. Mod. Phys.}\ }\textbf {\bibinfo {volume} {85}},\ \bibinfo
  {pages} {1143} (\bibinfo {year} {2013})}\BibitemShut {NoStop}%
\bibitem [{\citenamefont {Paxton}\ \emph {et~al.}(2004)\citenamefont {Paxton},
  \citenamefont {Kistler}, \citenamefont {Olmeda}, \citenamefont {Sen},
  \citenamefont {St.~Angelo}, \citenamefont {Cao}, \citenamefont {Mallouk},
  \citenamefont {Lammert},\ and\ \citenamefont {Crespi}}]{paxton2004catalytic}%
  \BibitemOpen
  \bibfield  {author} {\bibinfo {author} {\bibfnamefont {W.~F.}\ \bibnamefont
  {Paxton}}, \bibinfo {author} {\bibfnamefont {K.~C.}\ \bibnamefont {Kistler}},
  \bibinfo {author} {\bibfnamefont {C.~C.}\ \bibnamefont {Olmeda}}, \bibinfo
  {author} {\bibfnamefont {A.}~\bibnamefont {Sen}}, \bibinfo {author}
  {\bibfnamefont {S.~K.}\ \bibnamefont {St.~Angelo}}, \bibinfo {author}
  {\bibfnamefont {Y.}~\bibnamefont {Cao}}, \bibinfo {author} {\bibfnamefont
  {T.~E.}\ \bibnamefont {Mallouk}}, \bibinfo {author} {\bibfnamefont {P.~E.}\
  \bibnamefont {Lammert}}, \ and\ \bibinfo {author} {\bibfnamefont {V.~H.}\
  \bibnamefont {Crespi}},\ }\href@noop {} {\bibfield  {journal} {\bibinfo
  {journal} {J. Am. Chem. Soc}\ }\textbf {\bibinfo {volume} {126}},\ \bibinfo
  {pages} {13424} (\bibinfo {year} {2004})}\BibitemShut {NoStop}%
\bibitem [{\citenamefont {Dreyfus}\ \emph {et~al.}(2005)\citenamefont
  {Dreyfus}, \citenamefont {Baudry}, \citenamefont {Roper}, \citenamefont
  {Fermigier}, \citenamefont {Stone},\ and\ \citenamefont
  {Bibette}}]{dreyfus2005microscopic}%
  \BibitemOpen
  \bibfield  {author} {\bibinfo {author} {\bibfnamefont {R.}~\bibnamefont
  {Dreyfus}}, \bibinfo {author} {\bibfnamefont {J.}~\bibnamefont {Baudry}},
  \bibinfo {author} {\bibfnamefont {M.~L.}\ \bibnamefont {Roper}}, \bibinfo
  {author} {\bibfnamefont {M.}~\bibnamefont {Fermigier}}, \bibinfo {author}
  {\bibfnamefont {H.~A.}\ \bibnamefont {Stone}}, \ and\ \bibinfo {author}
  {\bibfnamefont {J.}~\bibnamefont {Bibette}},\ }\href@noop {} {\bibfield
  {journal} {\bibinfo  {journal} {Nature}\ }\textbf {\bibinfo {volume} {437}},\
  \bibinfo {pages} {862} (\bibinfo {year} {2005})}\BibitemShut {NoStop}%
\bibitem [{\citenamefont {Howse}\ \emph {et~al.}(2007)\citenamefont {Howse},
  \citenamefont {Jones}, \citenamefont {Ryan}, \citenamefont {Gough},
  \citenamefont {Vafabakhsh},\ and\ \citenamefont
  {Golestanian}}]{howse2007self}%
  \BibitemOpen
  \bibfield  {author} {\bibinfo {author} {\bibfnamefont {J.~R.}\ \bibnamefont
  {Howse}}, \bibinfo {author} {\bibfnamefont {R.~A.}\ \bibnamefont {Jones}},
  \bibinfo {author} {\bibfnamefont {A.~J.}\ \bibnamefont {Ryan}}, \bibinfo
  {author} {\bibfnamefont {T.}~\bibnamefont {Gough}}, \bibinfo {author}
  {\bibfnamefont {R.}~\bibnamefont {Vafabakhsh}}, \ and\ \bibinfo {author}
  {\bibfnamefont {R.}~\bibnamefont {Golestanian}},\ }\href@noop {} {\bibfield
  {journal} {\bibinfo  {journal} {Phys. Rev. Lett.}\ }\textbf {\bibinfo
  {volume} {99}},\ \bibinfo {pages} {048102} (\bibinfo {year}
  {2007})}\BibitemShut {NoStop}%
\bibitem [{\citenamefont {Hong}\ \emph {et~al.}(2007)\citenamefont {Hong},
  \citenamefont {Blackman}, \citenamefont {Kopp}, \citenamefont {Sen},\ and\
  \citenamefont {Velegol}}]{hong2007chemotaxis}%
  \BibitemOpen
  \bibfield  {author} {\bibinfo {author} {\bibfnamefont {Y.}~\bibnamefont
  {Hong}}, \bibinfo {author} {\bibfnamefont {N.~M.}\ \bibnamefont {Blackman}},
  \bibinfo {author} {\bibfnamefont {N.~D.}\ \bibnamefont {Kopp}}, \bibinfo
  {author} {\bibfnamefont {A.}~\bibnamefont {Sen}}, \ and\ \bibinfo {author}
  {\bibfnamefont {D.}~\bibnamefont {Velegol}},\ }\href@noop {} {\bibfield
  {journal} {\bibinfo  {journal} {Phys. Rev. Lett.}\ }\textbf {\bibinfo
  {volume} {99}},\ \bibinfo {pages} {178103} (\bibinfo {year}
  {2007})}\BibitemShut {NoStop}%
\bibitem [{\citenamefont {Deseigne}, \citenamefont {Dauchot},\ and\
  \citenamefont {Chat{\'e}}(2010)}]{deseigne2010collective}%
  \BibitemOpen
  \bibfield  {author} {\bibinfo {author} {\bibfnamefont {J.}~\bibnamefont
  {Deseigne}}, \bibinfo {author} {\bibfnamefont {O.}~\bibnamefont {Dauchot}}, \
  and\ \bibinfo {author} {\bibfnamefont {H.}~\bibnamefont {Chat{\'e}}},\
  }\href@noop {} {\bibfield  {journal} {\bibinfo  {journal} {Phys. Rev. Lett.}\
  }\textbf {\bibinfo {volume} {105}},\ \bibinfo {pages} {098001} (\bibinfo
  {year} {2010})}\BibitemShut {NoStop}%
\bibitem [{\citenamefont {Kudrolli}\ \emph {et~al.}(2008)\citenamefont
  {Kudrolli}, \citenamefont {Lumay}, \citenamefont {Volfson},\ and\
  \citenamefont {Tsimring}}]{kudrolli2008swarming}%
  \BibitemOpen
  \bibfield  {author} {\bibinfo {author} {\bibfnamefont {A.}~\bibnamefont
  {Kudrolli}}, \bibinfo {author} {\bibfnamefont {G.}~\bibnamefont {Lumay}},
  \bibinfo {author} {\bibfnamefont {D.}~\bibnamefont {Volfson}}, \ and\
  \bibinfo {author} {\bibfnamefont {L.~S.}\ \bibnamefont {Tsimring}},\ }\href
  {\doibase 10.1103/PhysRevLett.100.058001} {\bibfield  {journal} {\bibinfo
  {journal} {Phys. Rev. Lett.}\ }\textbf {\bibinfo {volume} {100}},\ \bibinfo
  {pages} {058001} (\bibinfo {year} {2008})}\BibitemShut {NoStop}%
\bibitem [{\citenamefont {Schaller}\ \emph {et~al.}(2010)\citenamefont
  {Schaller}, \citenamefont {Weber}, \citenamefont {Semmrich}, \citenamefont
  {Frey},\ and\ \citenamefont {Bausch}}]{schaller2010polar}%
  \BibitemOpen
  \bibfield  {author} {\bibinfo {author} {\bibfnamefont {V.}~\bibnamefont
  {Schaller}}, \bibinfo {author} {\bibfnamefont {C.}~\bibnamefont {Weber}},
  \bibinfo {author} {\bibfnamefont {C.}~\bibnamefont {Semmrich}}, \bibinfo
  {author} {\bibfnamefont {E.}~\bibnamefont {Frey}}, \ and\ \bibinfo {author}
  {\bibfnamefont {A.~R.}\ \bibnamefont {Bausch}},\ }\href@noop {} {\bibfield
  {journal} {\bibinfo  {journal} {Nature}\ }\textbf {\bibinfo {volume} {467}},\
  \bibinfo {pages} {73} (\bibinfo {year} {2010})}\BibitemShut {NoStop}%
\bibitem [{\citenamefont {Volpe}\ \emph {et~al.}(2011)\citenamefont {Volpe},
  \citenamefont {Buttinoni}, \citenamefont {Vogt}, \citenamefont
  {K{\"u}mmerer},\ and\ \citenamefont {Bechinger}}]{volpe2011microswimmers}%
  \BibitemOpen
  \bibfield  {author} {\bibinfo {author} {\bibfnamefont {G.}~\bibnamefont
  {Volpe}}, \bibinfo {author} {\bibfnamefont {I.}~\bibnamefont {Buttinoni}},
  \bibinfo {author} {\bibfnamefont {D.}~\bibnamefont {Vogt}}, \bibinfo {author}
  {\bibfnamefont {H.-J.}\ \bibnamefont {K{\"u}mmerer}}, \ and\ \bibinfo
  {author} {\bibfnamefont {C.}~\bibnamefont {Bechinger}},\ }\href@noop {}
  {\bibfield  {journal} {\bibinfo  {journal} {Soft Matter}\ }\textbf {\bibinfo
  {volume} {7}},\ \bibinfo {pages} {8810} (\bibinfo {year} {2011})}\BibitemShut
  {NoStop}%
\bibitem [{\citenamefont {Palacci}\ \emph {et~al.}(2013)\citenamefont
  {Palacci}, \citenamefont {Sacanna}, \citenamefont {Steinberg}, \citenamefont
  {Pine},\ and\ \citenamefont {Chaikin}}]{palacci2013living}%
  \BibitemOpen
  \bibfield  {author} {\bibinfo {author} {\bibfnamefont {J.}~\bibnamefont
  {Palacci}}, \bibinfo {author} {\bibfnamefont {S.}~\bibnamefont {Sacanna}},
  \bibinfo {author} {\bibfnamefont {A.~P.}\ \bibnamefont {Steinberg}}, \bibinfo
  {author} {\bibfnamefont {D.~J.}\ \bibnamefont {Pine}}, \ and\ \bibinfo
  {author} {\bibfnamefont {P.~M.}\ \bibnamefont {Chaikin}},\ }\href {\doibase
  10.1126/science.1230020} {\bibfield  {journal} {\bibinfo  {journal}
  {Science}\ }\textbf {\bibinfo {volume} {339}},\ \bibinfo {pages} {936}
  (\bibinfo {year} {2013})}\BibitemShut {NoStop}%
\bibitem [{\citenamefont {Bricard}\ \emph {et~al.}(2013)\citenamefont
  {Bricard}, \citenamefont {Caussin}, \citenamefont {Desreumaux}, \citenamefont
  {Dauchot},\ and\ \citenamefont {Bartolo}}]{bricard2013emergence}%
  \BibitemOpen
  \bibfield  {author} {\bibinfo {author} {\bibfnamefont {A.}~\bibnamefont
  {Bricard}}, \bibinfo {author} {\bibfnamefont {J.-B.}\ \bibnamefont
  {Caussin}}, \bibinfo {author} {\bibfnamefont {N.}~\bibnamefont {Desreumaux}},
  \bibinfo {author} {\bibfnamefont {O.}~\bibnamefont {Dauchot}}, \ and\
  \bibinfo {author} {\bibfnamefont {D.}~\bibnamefont {Bartolo}},\ }\href@noop
  {} {\bibfield  {journal} {\bibinfo  {journal} {Nature}\ }\textbf {\bibinfo
  {volume} {503}},\ \bibinfo {pages} {95} (\bibinfo {year} {2013})}\BibitemShut
  {NoStop}%
\bibitem [{\citenamefont {Wittkowski}\ \emph {et~al.}(2014)\citenamefont
  {Wittkowski}, \citenamefont {Tiribocchi}, \citenamefont {Stenhammar},
  \citenamefont {Allen}, \citenamefont {Marenduzzo},\ and\ \citenamefont
  {Cates}}]{wittkowski2014scalar}%
  \BibitemOpen
  \bibfield  {author} {\bibinfo {author} {\bibfnamefont {R.}~\bibnamefont
  {Wittkowski}}, \bibinfo {author} {\bibfnamefont {A.}~\bibnamefont
  {Tiribocchi}}, \bibinfo {author} {\bibfnamefont {J.}~\bibnamefont
  {Stenhammar}}, \bibinfo {author} {\bibfnamefont {R.~J.}\ \bibnamefont
  {Allen}}, \bibinfo {author} {\bibfnamefont {D.}~\bibnamefont {Marenduzzo}}, \
  and\ \bibinfo {author} {\bibfnamefont {M.~E.}\ \bibnamefont {Cates}},\
  }\href@noop {} {\bibfield  {journal} {\bibinfo  {journal} {‎Nat. Commun.}\
  }\textbf {\bibinfo {volume} {5}},\ \bibinfo {pages} {4531} (\bibinfo {year}
  {2014})}\BibitemShut {NoStop}%
\bibitem [{\citenamefont {Takatori}\ and\ \citenamefont
  {Brady}(2015)}]{takatori2015towards}%
  \BibitemOpen
  \bibfield  {author} {\bibinfo {author} {\bibfnamefont {S.~C.}\ \bibnamefont
  {Takatori}}\ and\ \bibinfo {author} {\bibfnamefont {J.~F.}\ \bibnamefont
  {Brady}},\ }\href@noop {} {\bibfield  {journal} {\bibinfo  {journal} {Phys.
  Rev. E}\ }\textbf {\bibinfo {volume} {91}},\ \bibinfo {pages} {032117}
  (\bibinfo {year} {2015})}\BibitemShut {NoStop}%
\bibitem [{\citenamefont {Maggi}\ \emph {et~al.}(2015)\citenamefont {Maggi},
  \citenamefont {Marconi}, \citenamefont {Gnan},\ and\ \citenamefont
  {Di~Leonardo}}]{maggi2015multidimensional}%
  \BibitemOpen
  \bibfield  {author} {\bibinfo {author} {\bibfnamefont {C.}~\bibnamefont
  {Maggi}}, \bibinfo {author} {\bibfnamefont {U.~M.~B.}\ \bibnamefont
  {Marconi}}, \bibinfo {author} {\bibfnamefont {N.}~\bibnamefont {Gnan}}, \
  and\ \bibinfo {author} {\bibfnamefont {R.}~\bibnamefont {Di~Leonardo}},\
  }\href@noop {} {\bibfield  {journal} {\bibinfo  {journal} {Sci. Rep.}\
  }\textbf {\bibinfo {volume} {5}},\ \bibinfo {pages} {10742} (\bibinfo {year}
  {2015})}\BibitemShut {NoStop}%
\bibitem [{\citenamefont {Solon}\ \emph {et~al.}(2015)\citenamefont {Solon},
  \citenamefont {Fily}, \citenamefont {Baskaran}, \citenamefont {Cates},
  \citenamefont {Kafri}, \citenamefont {Kardar},\ and\ \citenamefont
  {Tailleur}}]{solon2015pressure}%
  \BibitemOpen
  \bibfield  {author} {\bibinfo {author} {\bibfnamefont {A.}~\bibnamefont
  {Solon}}, \bibinfo {author} {\bibfnamefont {Y.}~\bibnamefont {Fily}},
  \bibinfo {author} {\bibfnamefont {A.}~\bibnamefont {Baskaran}}, \bibinfo
  {author} {\bibfnamefont {M.}~\bibnamefont {Cates}}, \bibinfo {author}
  {\bibfnamefont {Y.}~\bibnamefont {Kafri}}, \bibinfo {author} {\bibfnamefont
  {M.}~\bibnamefont {Kardar}}, \ and\ \bibinfo {author} {\bibfnamefont
  {J.}~\bibnamefont {Tailleur}},\ }\href@noop {} {\bibfield  {journal}
  {\bibinfo  {journal} {Nat. Phys.}\ }\textbf {\bibinfo {volume} {11}},\
  \bibinfo {pages} {673} (\bibinfo {year} {2015})}\BibitemShut {NoStop}%
\bibitem [{\citenamefont {Bialk{\'e}}\ \emph {et~al.}(2015)\citenamefont
  {Bialk{\'e}}, \citenamefont {Siebert}, \citenamefont {L{\"o}wen},\ and\
  \citenamefont {Speck}}]{bialke2015negative}%
  \BibitemOpen
  \bibfield  {author} {\bibinfo {author} {\bibfnamefont {J.}~\bibnamefont
  {Bialk{\'e}}}, \bibinfo {author} {\bibfnamefont {J.~T.}\ \bibnamefont
  {Siebert}}, \bibinfo {author} {\bibfnamefont {H.}~\bibnamefont {L{\"o}wen}},
  \ and\ \bibinfo {author} {\bibfnamefont {T.}~\bibnamefont {Speck}},\
  }\href@noop {} {\bibfield  {journal} {\bibinfo  {journal} {Phys. Rev. Lett.}\
  }\textbf {\bibinfo {volume} {115}},\ \bibinfo {pages} {098301} (\bibinfo
  {year} {2015})}\BibitemShut {NoStop}%
\bibitem [{\citenamefont {Falasco}\ \emph {et~al.}(2015)\citenamefont
  {Falasco}, \citenamefont {Baldovin}, \citenamefont {Kroy},\ and\
  \citenamefont {Baiesi}}]{falasco2015mesoscopic}%
  \BibitemOpen
  \bibfield  {author} {\bibinfo {author} {\bibfnamefont {G.}~\bibnamefont
  {Falasco}}, \bibinfo {author} {\bibfnamefont {F.}~\bibnamefont {Baldovin}},
  \bibinfo {author} {\bibfnamefont {K.}~\bibnamefont {Kroy}}, \ and\ \bibinfo
  {author} {\bibfnamefont {M.}~\bibnamefont {Baiesi}},\ }\href@noop {}
  {\bibfield  {journal} {\bibinfo  {journal} {arXiv:1512.01687}\ } (\bibinfo
  {year} {2015})}\BibitemShut {NoStop}%
\bibitem [{\citenamefont {Speck}\ and\ \citenamefont
  {Jack}(2016)}]{speck2015ideal}%
  \BibitemOpen
  \bibfield  {author} {\bibinfo {author} {\bibfnamefont {T.}~\bibnamefont
  {Speck}}\ and\ \bibinfo {author} {\bibfnamefont {R.~L.}\ \bibnamefont
  {Jack}},\ }\href@noop {} {\bibfield  {journal} {\bibinfo  {journal} {Phys.
  Rev. E}\ }\textbf {\bibinfo {volume} {93}},\ \bibinfo {pages} {062605}
  (\bibinfo {year} {2016})}\BibitemShut {NoStop}%
\bibitem [{\citenamefont {Speck}(2016)}]{speck2016stochastic}%
  \BibitemOpen
  \bibfield  {author} {\bibinfo {author} {\bibfnamefont {T.}~\bibnamefont
  {Speck}},\ }\href@noop {} {\bibfield  {journal} {\bibinfo  {journal} {EPL
  (Europhysics Letters)}\ }\textbf {\bibinfo {volume} {114}},\ \bibinfo {pages}
  {30006} (\bibinfo {year} {2016})}\BibitemShut {NoStop}%
\bibitem [{\citenamefont {Vicsek}\ \emph {et~al.}(1995)\citenamefont {Vicsek},
  \citenamefont {Czir\'ok}, \citenamefont {Ben-Jacob}, \citenamefont {Cohen},\
  and\ \citenamefont {Shochet}}]{vicsek1995novel}%
  \BibitemOpen
  \bibfield  {author} {\bibinfo {author} {\bibfnamefont {T.}~\bibnamefont
  {Vicsek}}, \bibinfo {author} {\bibfnamefont {A.}~\bibnamefont {Czir\'ok}},
  \bibinfo {author} {\bibfnamefont {E.}~\bibnamefont {Ben-Jacob}}, \bibinfo
  {author} {\bibfnamefont {I.}~\bibnamefont {Cohen}}, \ and\ \bibinfo {author}
  {\bibfnamefont {O.}~\bibnamefont {Shochet}},\ }\href@noop {} {\bibfield
  {journal} {\bibinfo  {journal} {Phys. Rev. Lett.}\ }\textbf {\bibinfo
  {volume} {75}},\ \bibinfo {pages} {1226} (\bibinfo {year}
  {1995})}\BibitemShut {NoStop}%
\bibitem [{\citenamefont {Chat{\'e}}, \citenamefont {Ginelli},\ and\
  \citenamefont {Montagne}(2006)}]{chate2006simple}%
  \BibitemOpen
  \bibfield  {author} {\bibinfo {author} {\bibfnamefont {H.}~\bibnamefont
  {Chat{\'e}}}, \bibinfo {author} {\bibfnamefont {F.}~\bibnamefont {Ginelli}},
  \ and\ \bibinfo {author} {\bibfnamefont {R.}~\bibnamefont {Montagne}},\
  }\href@noop {} {\bibfield  {journal} {\bibinfo  {journal} {Phys. Rev. Lett.}\
  }\textbf {\bibinfo {volume} {96}},\ \bibinfo {pages} {180602} (\bibinfo
  {year} {2006})}\BibitemShut {NoStop}%
\bibitem [{\citenamefont {Chat{\'e}}\ \emph {et~al.}(2008)\citenamefont
  {Chat{\'e}}, \citenamefont {Ginelli}, \citenamefont {Gr{\'e}goire},\ and\
  \citenamefont {Raynaud}}]{chate2008collective}%
  \BibitemOpen
  \bibfield  {author} {\bibinfo {author} {\bibfnamefont {H.}~\bibnamefont
  {Chat{\'e}}}, \bibinfo {author} {\bibfnamefont {F.}~\bibnamefont {Ginelli}},
  \bibinfo {author} {\bibfnamefont {G.}~\bibnamefont {Gr{\'e}goire}}, \ and\
  \bibinfo {author} {\bibfnamefont {F.}~\bibnamefont {Raynaud}},\ }\href@noop
  {} {\bibfield  {journal} {\bibinfo  {journal} {Phys. Rev. E}\ }\textbf
  {\bibinfo {volume} {77}},\ \bibinfo {pages} {046113} (\bibinfo {year}
  {2008})}\BibitemShut {NoStop}%
\bibitem [{\citenamefont {Ginelli}\ and\ \citenamefont
  {Chat{\'e}}(2010)}]{ginelli2010relevance}%
  \BibitemOpen
  \bibfield  {author} {\bibinfo {author} {\bibfnamefont {F.}~\bibnamefont
  {Ginelli}}\ and\ \bibinfo {author} {\bibfnamefont {H.}~\bibnamefont
  {Chat{\'e}}},\ }\href@noop {} {\bibfield  {journal} {\bibinfo  {journal}
  {Phys. Rev. Lett.}\ }\textbf {\bibinfo {volume} {105}},\ \bibinfo {pages}
  {168103} (\bibinfo {year} {2010})}\BibitemShut {NoStop}%
\bibitem [{\citenamefont {Dey}, \citenamefont {Das},\ and\ \citenamefont
  {Rajesh}(2012)}]{dey2012spatial}%
  \BibitemOpen
  \bibfield  {author} {\bibinfo {author} {\bibfnamefont {S.}~\bibnamefont
  {Dey}}, \bibinfo {author} {\bibfnamefont {D.}~\bibnamefont {Das}}, \ and\
  \bibinfo {author} {\bibfnamefont {R.}~\bibnamefont {Rajesh}},\ }\href@noop {}
  {\bibfield  {journal} {\bibinfo  {journal} {Phys. Rev. Lett.}\ }\textbf
  {\bibinfo {volume} {108}},\ \bibinfo {pages} {238001} (\bibinfo {year}
  {2012})}\BibitemShut {NoStop}%
\bibitem [{\citenamefont {Bechinger}\ \emph {et~al.}(2016)\citenamefont
  {Bechinger}, \citenamefont {Di~Leonardo}, \citenamefont {L{\"o}wen},
  \citenamefont {Reichhardt}, \citenamefont {Volpe},\ and\ \citenamefont
  {Volpe}}]{bechinger2016active}%
  \BibitemOpen
  \bibfield  {author} {\bibinfo {author} {\bibfnamefont {C.}~\bibnamefont
  {Bechinger}}, \bibinfo {author} {\bibfnamefont {R.}~\bibnamefont
  {Di~Leonardo}}, \bibinfo {author} {\bibfnamefont {H.}~\bibnamefont
  {L{\"o}wen}}, \bibinfo {author} {\bibfnamefont {C.}~\bibnamefont
  {Reichhardt}}, \bibinfo {author} {\bibfnamefont {G.}~\bibnamefont {Volpe}}, \
  and\ \bibinfo {author} {\bibfnamefont {G.}~\bibnamefont {Volpe}},\
  }\href@noop {} {\bibfield  {journal} {\bibinfo  {journal} {arXiv:1602.00081}\
  } (\bibinfo {year} {2016})}\BibitemShut {NoStop}%
\bibitem [{\citenamefont {Smit}(1992)}]{smit1992phase}%
  \BibitemOpen
  \bibfield  {author} {\bibinfo {author} {\bibfnamefont {B.}~\bibnamefont
  {Smit}},\ }\href@noop {} {\bibfield  {journal} {\bibinfo  {journal} {J. Chem.
  Phys.}\ }\textbf {\bibinfo {volume} {96}},\ \bibinfo {pages} {8639} (\bibinfo
  {year} {1992})}\BibitemShut {NoStop}%
\bibitem [{\citenamefont {Vrabec}\ \emph {et~al.}(2006)\citenamefont {Vrabec},
  \citenamefont {Kedia}, \citenamefont {Fuchs},\ and\ \citenamefont
  {Hasse}}]{vrabec2006comprehensive}%
  \BibitemOpen
  \bibfield  {author} {\bibinfo {author} {\bibfnamefont {J.}~\bibnamefont
  {Vrabec}}, \bibinfo {author} {\bibfnamefont {G.~K.}\ \bibnamefont {Kedia}},
  \bibinfo {author} {\bibfnamefont {G.}~\bibnamefont {Fuchs}}, \ and\ \bibinfo
  {author} {\bibfnamefont {H.}~\bibnamefont {Hasse}},\ }\href@noop {}
  {\bibfield  {journal} {\bibinfo  {journal} {Mol. Phys.}\ }\textbf {\bibinfo
  {volume} {104}},\ \bibinfo {pages} {1509} (\bibinfo {year}
  {2006})}\BibitemShut {NoStop}%
\bibitem [{\citenamefont {Yang}\ and\ \citenamefont
  {Lee}(1952)}]{yang1952statistical}%
  \BibitemOpen
  \bibfield  {author} {\bibinfo {author} {\bibfnamefont {C.-N.}\ \bibnamefont
  {Yang}}\ and\ \bibinfo {author} {\bibfnamefont {T.-D.}\ \bibnamefont {Lee}},\
  }\href@noop {} {\bibfield  {journal} {\bibinfo  {journal} {Phys. Rev.}\
  }\textbf {\bibinfo {volume} {87}},\ \bibinfo {pages} {404} (\bibinfo {year}
  {1952})}\BibitemShut {NoStop}%
\bibitem [{\citenamefont {Lee}\ and\ \citenamefont
  {Yang}(1952)}]{lee1952statistical2}%
  \BibitemOpen
  \bibfield  {author} {\bibinfo {author} {\bibfnamefont {T.-D.}\ \bibnamefont
  {Lee}}\ and\ \bibinfo {author} {\bibfnamefont {C.-N.}\ \bibnamefont {Yang}},\
  }\href@noop {} {\bibfield  {journal} {\bibinfo  {journal} {Phys. Rev.}\
  }\textbf {\bibinfo {volume} {87}},\ \bibinfo {pages} {410} (\bibinfo {year}
  {1952})}\BibitemShut {NoStop}%
\bibitem [{\citenamefont {Watanabe}, \citenamefont {Ito},\ and\ \citenamefont
  {Hu}(2012)}]{watanabe2012phase}%
  \BibitemOpen
  \bibfield  {author} {\bibinfo {author} {\bibfnamefont {H.}~\bibnamefont
  {Watanabe}}, \bibinfo {author} {\bibfnamefont {N.}~\bibnamefont {Ito}}, \
  and\ \bibinfo {author} {\bibfnamefont {C.-K.}\ \bibnamefont {Hu}},\
  }\href@noop {} {\bibfield  {journal} {\bibinfo  {journal} {J. Chem. Phys.}\
  }\textbf {\bibinfo {volume} {136}},\ \bibinfo {pages} {204102} (\bibinfo
  {year} {2012})}\BibitemShut {NoStop}%
\bibitem [{\citenamefont {Reif-Acherman}(2010)}]{reif2010history}%
  \BibitemOpen
  \bibfield  {author} {\bibinfo {author} {\bibfnamefont {S.}~\bibnamefont
  {Reif-Acherman}},\ }\href@noop {} {\bibfield  {journal} {\bibinfo  {journal}
  {Qu\'i\-m. Nova}\ }\textbf {\bibinfo {volume} {33}},\ \bibinfo {pages} {2003}
  (\bibinfo {year} {2010})}\BibitemShut {NoStop}%
\bibitem [{\citenamefont {Prymidis}, \citenamefont {Samin},\ and\ \citenamefont
  {Filion}(2016)}]{prymidis2016state}%
  \BibitemOpen
  \bibfield  {author} {\bibinfo {author} {\bibfnamefont {V.}~\bibnamefont
  {Prymidis}}, \bibinfo {author} {\bibfnamefont {S.}~\bibnamefont {Samin}}, \
  and\ \bibinfo {author} {\bibfnamefont {L.}~\bibnamefont {Filion}},\
  }\href@noop {} {\bibfield  {journal} {\bibinfo  {journal} {Soft matter}\
  }\textbf {\bibinfo {volume} {12}},\ \bibinfo {pages} {4309} (\bibinfo {year}
  {2016})}\BibitemShut {NoStop}%
\bibitem [{\citenamefont {Noro}\ and\ \citenamefont
  {Frenkel}(2000)}]{noro2000extended}%
  \BibitemOpen
  \bibfield  {author} {\bibinfo {author} {\bibfnamefont {M.~G.}\ \bibnamefont
  {Noro}}\ and\ \bibinfo {author} {\bibfnamefont {D.}~\bibnamefont {Frenkel}},\
  }\href@noop {} {\bibfield  {journal} {\bibinfo  {journal} {J. Chem. Phys.}\
  }\textbf {\bibinfo {volume} {113}},\ \bibinfo {pages} {2941} (\bibinfo {year}
  {2000})}\BibitemShut {NoStop}%
\bibitem [{\citenamefont {Dunikov}, \citenamefont {Malyshenko},\ and\
  \citenamefont {Zhakhovskii}(2001)}]{dunikov2001corresponding}%
  \BibitemOpen
  \bibfield  {author} {\bibinfo {author} {\bibfnamefont {D.}~\bibnamefont
  {Dunikov}}, \bibinfo {author} {\bibfnamefont {S.}~\bibnamefont {Malyshenko}},
  \ and\ \bibinfo {author} {\bibfnamefont {V.}~\bibnamefont {Zhakhovskii}},\
  }\href@noop {} {\bibfield  {journal} {\bibinfo  {journal} {J. Chem. Phys.}\
  }\textbf {\bibinfo {volume} {115}},\ \bibinfo {pages} {6623} (\bibinfo {year}
  {2001})}\BibitemShut {NoStop}%
\bibitem [{\citenamefont {Redner}, \citenamefont {Baskaran},\ and\
  \citenamefont {Hagan}(2013)}]{redner2013reentrant}%
  \BibitemOpen
  \bibfield  {author} {\bibinfo {author} {\bibfnamefont {G.~S.}\ \bibnamefont
  {Redner}}, \bibinfo {author} {\bibfnamefont {A.}~\bibnamefont {Baskaran}}, \
  and\ \bibinfo {author} {\bibfnamefont {M.~F.}\ \bibnamefont {Hagan}},\
  }\href@noop {} {\bibfield  {journal} {\bibinfo  {journal} {Phys. Rev. E}\
  }\textbf {\bibinfo {volume} {88}},\ \bibinfo {pages} {012305} (\bibinfo
  {year} {2013})}\BibitemShut {NoStop}%
\bibitem [{\citenamefont {Mognetti}\ \emph {et~al.}(2013)\citenamefont
  {Mognetti}, \citenamefont {\ifmmode \check{S}\else
  \v{S}\fi{}ari\ifmmode~\acute{c}\else \'{c}\fi{}}, \citenamefont
  {Angioletti-Uberti}, \citenamefont {Cacciuto}, \citenamefont {Valeriani},\
  and\ \citenamefont {Frenkel}}]{Mognetti2013}%
  \BibitemOpen
  \bibfield  {author} {\bibinfo {author} {\bibfnamefont {B.~M.}\ \bibnamefont
  {Mognetti}}, \bibinfo {author} {\bibfnamefont {A.}~\bibnamefont {\ifmmode
  \check{S}\else \v{S}\fi{}ari\ifmmode~\acute{c}\else \'{c}\fi{}}}, \bibinfo
  {author} {\bibfnamefont {S.}~\bibnamefont {Angioletti-Uberti}}, \bibinfo
  {author} {\bibfnamefont {A.}~\bibnamefont {Cacciuto}}, \bibinfo {author}
  {\bibfnamefont {C.}~\bibnamefont {Valeriani}}, \ and\ \bibinfo {author}
  {\bibfnamefont {D.}~\bibnamefont {Frenkel}},\ }\href@noop {} {\bibfield
  {journal} {\bibinfo  {journal} {Phys. Rev. Lett.}\ }\textbf {\bibinfo
  {volume} {111}},\ \bibinfo {pages} {245702} (\bibinfo {year}
  {2013})}\BibitemShut {NoStop}%
\bibitem [{\citenamefont {Prymidis}, \citenamefont {Sielcken},\ and\
  \citenamefont {Filion}(2015)}]{prymidis2015self}%
  \BibitemOpen
  \bibfield  {author} {\bibinfo {author} {\bibfnamefont {V.}~\bibnamefont
  {Prymidis}}, \bibinfo {author} {\bibfnamefont {H.}~\bibnamefont {Sielcken}},
  \ and\ \bibinfo {author} {\bibfnamefont {L.}~\bibnamefont {Filion}},\
  }\href@noop {} {\bibfield  {journal} {\bibinfo  {journal} {Soft Matter}\
  }\textbf {\bibinfo {volume} {11}},\ \bibinfo {pages} {4158} (\bibinfo {year}
  {2015})}\BibitemShut {NoStop}%
\bibitem [{\citenamefont {Higham}(2001)}]{Higham2001}%
  \BibitemOpen
  \bibfield  {author} {\bibinfo {author} {\bibfnamefont {D.}~\bibnamefont
  {Higham}},\ }\href {\doibase 10.1137/S0036144500378302} {\bibfield  {journal}
  {\bibinfo  {journal} {SIAM Rev.}\ }\textbf {\bibinfo {volume} {43}},\
  \bibinfo {pages} {525} (\bibinfo {year} {2001})}\BibitemShut {NoStop}%
\bibitem [{\citenamefont {Winkler}, \citenamefont {Wysocki},\ and\
  \citenamefont {Gompper}(2015)}]{winkler2015virial}%
  \BibitemOpen
  \bibfield  {author} {\bibinfo {author} {\bibfnamefont {R.~G.}\ \bibnamefont
  {Winkler}}, \bibinfo {author} {\bibfnamefont {A.}~\bibnamefont {Wysocki}}, \
  and\ \bibinfo {author} {\bibfnamefont {G.}~\bibnamefont {Gompper}},\
  }\href@noop {} {\bibfield  {journal} {\bibinfo  {journal} {Soft matter}\
  }\textbf {\bibinfo {volume} {11}},\ \bibinfo {pages} {6680} (\bibinfo {year}
  {2015})}\BibitemShut {NoStop}%
\bibitem [{\citenamefont {Takatori}, \citenamefont {Yan},\ and\ \citenamefont
  {Brady}(2014)}]{takatori2014swim}%
  \BibitemOpen
  \bibfield  {author} {\bibinfo {author} {\bibfnamefont {S.~C.}\ \bibnamefont
  {Takatori}}, \bibinfo {author} {\bibfnamefont {W.}~\bibnamefont {Yan}}, \
  and\ \bibinfo {author} {\bibfnamefont {J.~F.}\ \bibnamefont {Brady}},\
  }\href@noop {} {\bibfield  {journal} {\bibinfo  {journal} {Phys. Rev. Lett.}\
  }\textbf {\bibinfo {volume} {113}},\ \bibinfo {pages} {028103} (\bibinfo
  {year} {2014})}\BibitemShut {NoStop}%
\bibitem [{\citenamefont {Guggenheim}(1945)}]{guggenheim1945principle}%
  \BibitemOpen
  \bibfield  {author} {\bibinfo {author} {\bibfnamefont {E.~A.}\ \bibnamefont
  {Guggenheim}},\ }\href@noop {} {\bibfield  {journal} {\bibinfo  {journal} {J.
  Chem. Phys.}\ }\textbf {\bibinfo {volume} {13}},\ \bibinfo {pages} {253}
  (\bibinfo {year} {1945})}\BibitemShut {NoStop}%
\end{thebibliography}%

\end{document}